\documentclass[twocolumn]{el-author}

\usepackage{cite}
\usepackage{amsmath,amssymb,amsfonts,amsthm}
\usepackage{stmaryrd}
\usepackage{algorithmic}
\usepackage{graphicx}
\usepackage{textcomp}
\usepackage{xcolor}
\usepackage{makecell}
\usepackage{optidef}
\usepackage{algorithm}
\usepackage{algorithmic}
\usepackage{adjustbox}
\usepackage{flushend}

\usepackage{subfig}
\usepackage[subtle]{savetrees}

\newcommand{\ua}{\uparrow}
\newcommand{\nc}{\newcommand}
\nc{\da}{\downarrow} \nc{\hc}{\hat{c}} \nc{\hS}{\hat{S}}
\nc{\bra}{\langle} \nc{\ket}{\rangle} \nc{\eq}{equation (\ref}
\nc{\h}{\hat} \nc{\hT}{\h{T}}\nc{\be}{\begin{eqnarray}}
\nc{\ee}{\end{eqnarray}}\nc{\rd}{\textrm{d}}\nc{\e}{eqnarray}\nc{\hR}{\hat{R}}\nc{\Tr}{\mathrm{Tr}}
\nc{\tS}{\tilde{S}}\nc{\tr}{\mathrm{tr}}\nc{\8}{\infty}\nc{\lgs}{\bra\ua,\phi|}\nc{\rgs}{|\ua,\phi\ket}
\nc{\hU}{\hat{U}}\nc{\lfs}{\bra\phi|}\nc{\rfs}{|\phi\ket}\nc{\hZ}{\hat{Z}}\nc{\hd}{\hat{d}}\nc{\mD}{\mathcal{D}}
\nc{\bd}{\bar{d}}\nc{\bc}{\bar{c}}\nc{\mc}{\mathcal}\nc{\ea}{eqnarray}\nc{\mG}{\mathcal{G}}\nc{\bce}{\begin{center}}
\nc{\ece}{\end{center}}
\begin{document}
\title{Active learning for efficient data selection in radio-signal based positioning via deep learning}
\author{V. Corlay and M. Courcoux-Caro}
	
\abstract{We consider the problem of user equipment (UE) positioning based on radio signals via deep learning.
As in most supervised-learning tasks, a critical aspect is the availability of a relevant dataset to train a model.
However, in a cellular network, the data-collection step may induce a high communication overhead. 
As a result, to reduce the required size of the dataset, it may be interesting to carefully choose the positions to be labelled and to be used in the training.
We therefore propose an active learning approach for efficient data collection. 
We first show that significant gains (both in terms of positioning accuracy and size of the required dataset) can be obtained for the considered positioning problem using a genie. This validates the interest of active learning for positioning. We then propose a practical method to approximate this genie.}

\maketitle


\section{Introduction}

Standard positioning algorithms based on radio signals, as the ones considered in 5G NR \cite{Garcia5GBOOK}, rely on line-of-sight (LoS) measurements.
More advanced algorithms need therefore to be considered for non line-of-sight (NLoS) cases where the conventional positioning techniques perform poorly \cite{TdocSumOct}.
This includes indoor situations as well as challenging urban environments.
One approach to address this challenge is to use a digital twin as proposed in e.g., \cite{Nguyen2023}\cite{Corlay2024}. This can be seen as a model-based approach.
Unsupervised learning approaches may also be considered, e.g., by pre-training offline a model with model-based data such that the online supervised training part can be performed with a low amount of data \cite{Arnold2018}.
Alternatively, if many data are available, supervised deep learning is a key technology to improve the positioning accuracy.
This approach is the one investigated in the technical report of the 3GPP on artificial intelligence for the NR air interface \cite{TdocSumOct}.
In this paper, we focus on this supervised learning approach. The proposed technique can also be used for data selection in a supervised fine-tuning step, where another training method can be used to pre-train the model (similarly to \cite{Arnold2018}).

A central aspect of the design of deep learning models trained in a supervised manner is data collection, i.e., obtain radio signals corresponding to known positions. 
However, in a cellular-network context, if the data to train the models is collected in an online manner, this may induce a high overhead on the communication network. For example, the purpose of \cite{Arnold2018} is also to reduce the amount of data to be collected in a radio-based positioning problem. More generally, it has been a long-term research goal in the field of wireless communication to reduce the pilot overhead \cite{Hassibi2000}.
A popular approach to reduce the overhead induced by the training of a supervised model is federated learning \cite{McMahan2017}. This was introduced in 2016 as an efficient decentralized solution, where the local datasets of the user equipments (UE) do not need to be uploaded to the central server. This solution was recently standardized by the 3GPP \cite[Sec. 6.2C]{TS23288}.
However, it cannot be used in our case: the fingerprint radio signals are not apriori possessed by the UE or the BS.
 
As a result, we believe that the following questions are relevant: When should the base stations (BS) collect the positions and the corresponding radio signals of the UE? Should it be done when the UE is at specific positions or completely randomly?

\textbf{Contributions.}
In order to reduce the amount of data required, we propose a new data-selection technique. The aim is to efficiently train a neural network in a supervised manner to infer the position of a UE based on a received radio signal. The technique enables to select (i.e., obtain the label of) the most relevant training positions to improve a model after a first training phase. As explained in the following ``literature review", this idea of selecting the data to be labelled during the training process belongs to the active learning paradigm. In this paper, we therefore propose a new active learning method.

We first provide simulation results showing that carefully choosing the data to train a model 
can significantly improve the performance compared to the random selection.
This validates that there is an interest to study data-selection techniques.
However, the proposed protocol to obtain this conclusion cannot be applied in practice as it uses information not available by the entity performing the selection.
We therefore call this approach the genie approach.
Hence, the performance obtained with the genie can be seen as an upper bound of the achievable performance.
We then propose a practical algorithm to try to approximate the genie. This algorithm exhibits better performance than a random selection.

More specifically, we investigate via numerical simulations the gains obtained by selecting $10\%$ of $N$ candidate positions with the proposed methods, to be compared with the random selection of these $10\%$. With the considered parameters, simulation results show that the genie approach enables to get more than $50\%$ of the achievable gain (the full achievable gain being computed using all $N$ candidate positions for the training), where the random approach achieves $15\%$. In terms of data savings, we obtain that the genie approach achieves similar performance as the random approach using approximately only 20$\%$ of the data.
Regarding the performance of the practical algorithm, it achieves approximately $50\%$ of the gain of the genie approach.

Finally, while the proposed data-selection method is used to address the positioning problem, note that it can be considered in any field where there is a strong constraint on the data collection.

\textbf{Literature review.}
As mentioned in the ``contributions", the proposed approach can be seen as a form of active learning \cite{Settles2009}\cite{Meng2017}\cite{Ren2021}. The main idea of active learning is the following. If the learning algorithm can choose the data from which it learns, it will perform better with less training.  More specifically, active learning performs queries of unlabelled instances that should be labelled.
The goal of an active learner is to achieve high accuracy using as few labelled instances as possible.  

Deep learning for  radio-based positioning has been widely studied in the literature \cite{TdocSumOct}\cite{Arnold2018}\cite{Chatelier2023}\cite{GutoLeoni2020}\cite{Ferenc2022}.
Several studies such as \cite{TdocSumOct}\cite{Chatelier2023} report the positioning performance as a function of the dataset size. 
The reference \cite{TdocSumOct} is the recent first version
of a 3GPP technical report focusing on “Study on Artificial intelligence (AI)/Machine Learning (ML) for NR air interface”, where one of the three main use cases considered is the NLoS positioning problem. 
Nevertheless, to the best of the authors knowledge, no active learning approach has been considered for efficient data selection in the scope of the positioning problem. 

\textbf{Considered problem.}
The goal is to train a model (i.e., a neural network) to infer the position of a UE based on radio signals measured at several BS\footnote{Equivalently, the UE can measure the downlink radio signals of several BS and report it to the network.}. As a result, the $i$-th dataset $\mathcal{D}_i$, used to train the model, is a set of many couples (Radio signals, position). The ``radio signals" include the measurements of all BS. We assume that the training is performed at the network side, where the dataset $\mathcal{D}_i$ is available.
The type of radio signal considered in this work is mainly the path gain (PG). The channel impulse response (CIR) is also considered for complementary simulations. 

We consider the following protocol.
We assume that an initial dataset $\mathcal{D}_1$ is available at the network side, and used to train a model called Neural Network 1A (NN1A) to infer the position. This results in a given performance level of NN1A. Then, we have $N$ new candidate positions and we want to select only X$\%$ of these positions (e.g., $X\%=10\%$). The initial dataset $\mathcal{D}_1$ and the parameter X are assumed to be fixed (system constrains).
The selected positions and the corresponding radio signals are added in $\mathcal{D}_1$ to obtain the dataset $\mathcal{D}_2$. 
Finally, the network NN1A is then further trained with $\mathcal{D}_2$. 
Note that the first step (initial training of NN1A) could also be realized by any training approach (e.g., unsupervised)  such that the data-selection technique is used only for a supervised fine-tuning step (similarly to \cite{Arnold2018} where supervised training with real data is used only for fine tuning).
We summarize the protocol with the following steps below:
\begin{enumerate}
\item Train NN1A with $\mathcal{D}_1$ (or with another training method). 
\item Select $X \%$ of $N$ new candidate positions. Three data-selection algorithms are considered:
\begin{itemize}
\item Random selection.
\item Genie selection.
\item Practical selection.
\end{itemize}
\item Add the selected positions in $\mathcal{D}_1$ to obtain $\mathcal{D}_2$.
\item Further train NN1A with $\mathcal{D}_2$ to obtain an improved version of NN1A called NN2A.
\end{enumerate}

The $N$ candidate positions correspond for instance to the possible locations where the UE could be in the future. This information can be established based on the ``expected UE behavior provision information" as referenced in the 3GPP technical report \cite{TR23700}.
Alternatively, these $N$ positions can be random positions or positions of a 2D grid covering the scene not included in $\mathcal{D}_1$.
The $X\%$ selected positions would then be the positions where the UE is configured to send an uplink radio signal such that the network can perform a measurement and thus collect the relevant data. Note that the selection of the candidate positions itself could be a topic of research. In this paper, we focus on the selection the best positions in a fixed (not optimized) candidate set. For the sake of simplicity, we take the size of $\mathcal{D}_1$ also equal to $N$ such that after adding the selected positions, $\mathcal{D}_2$ has a size $1.(X/100) \times N$.

The benchmark to assess the proposed algorithms (genie and practical) is the above-mentioned random selection: it consists in selecting randomly the X$\%$ positions among the $N$ positions to be added in $\mathcal{D}_1$.
Hence, this paper investigates whether there is an advantage to carefully select the additional positions over a random selection and how it could be performed. 


\section{Proposed approaches for data selection}
\label{sec_proposed_tech}
\textbf{Genie approach.}
We propose below a first protocol for data selection. This first protocol uses information that should not be available at the training location. 
We therefore call it the genie approach. As mentioned in the introduction, the goal here is to show that significant performance gain can be obtained over a random selection. 

The main idea consists in selecting, among the $N$ candidate positions, the positions on which NN1A (pre-trained with $\mathcal{D}_1$) performs the worst.
Intuitively, these positions should yield more improvement on the model than the positions already correctly predicted by the model.
In order to validate this intuition, we propose the following protocol (details of the above steps 2-4).

\begin{enumerate}
\item We assume that a genie has access to the radio signals corresponding to the $N$ candidate positions (this is not possible in a real situation).
\item We test NN1A (pre-trained with $\mathcal{D}_1$) on all these estimated radio signals and compute the errors between the obtained positions and the true positions.
\item We add the $X\%$ positions with the greatest error in the dataset $\mathcal{D}_1$ to obtain~$\mathcal{D}_2$.
\item The network NN1A is then further trained with $\mathcal{D}_2$ to obtain the neural network NN2A.
\end{enumerate}

The benchmark random data selection selects the positions to be added in $\mathcal{D}_1$ by sampling randomly $X\%$ of the $N$ positions. 
The simulation results with $X\%=10\%$, provided at the end of the paper, show that the genie approach offers a significant gain over the random data selection. This validates the interest to study practical data-selection algorithms. Such an algorithm is proposed in the following subsection.

\begin{figure*}[t]
    \centering
    \includegraphics[scale=0.5]{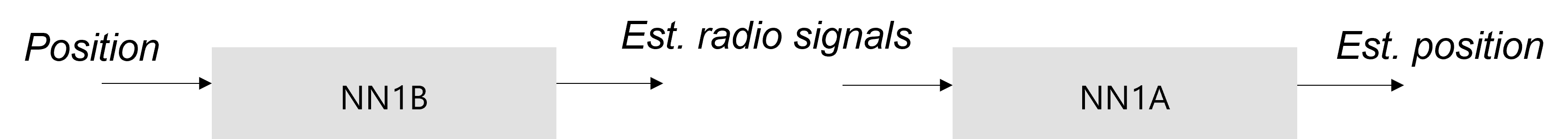}
    \caption{Proposed usage of NN1B and NN1A for efficient data selection. The goal of NN1B is to approximate the genie.}
    \label{fig:selection with NN1A and NN1B}
\end{figure*}
\textbf{Practical algorithm to approximate the genie approach.}
In this subsection, we propose an algorithm to approximate the genie.
As explained in the previous section, the radio signals corresponding the $N$ candidate positions are not available at the training location.
Hence, in addition to training the model NN1A with the dataset $\mathcal{D}_1$, we propose to train a second model, called NN1B.
This second model NN1B is trained with the same dataset $\mathcal{D}_1$, but the input and the label are inverted, i.e., it takes the position as input and outputs the radio signals. 
The goal of NN1B is to estimate the radio signals corresponding to each of the $N$ candidate positions (which is provided by the genie in the previous subsection). 
Then, the rest of the algorithm is unchanged: We test NN1A on all these (estimated) radio signals and compute the errors between the obtained positions and the true positions.  Fig.~\ref{fig:selection with NN1A and NN1B} illustrates the use of NN1B and NN1A for data selection. 
This yields an autoencoder-like structure for inference. 
Note however that there is no training in this data selection part. NN1B and NN1A remain unchanged.
The full practical algorithm is summarized below:
\begin{enumerate}
\item Train NN1A and NN1B with $\mathcal{D}_1$.
\item Select $X \%$ of $N$ candidate positions by using NN1A and NN1B as on Fig.~\ref{fig:selection with NN1A and NN1B}.
\item Add the selected positions in $\mathcal{D}_1$ to obtain $\mathcal{D}_2$.
\item Further train NN1A with $\mathcal{D}_2$ to obtain NN2A.
\end{enumerate}


\section{Considered datasets and neural network architectures}
\label{Sec_dataset}
We consider a dataset obtained via a statistical channel: the 3GPP InF-DH scenario as described in TR 38.901 \cite{TR38901}.
We recently used this dataset in \cite{Chatelier2023} to investigate the influence of some radio-signal dataset parameters on the positioning performance.
This statistical channel is also used by the 3GPP in the study on NLoS positioning with deep learning \cite{TdocSumOct}.
The topology of the factory  in the 3GPP InF model is a rectangular area of width $W=60$m, length $L=120$m, and height $H=10$m. There are up to 18 BS at specific locations with inter-BS spacing $D=20$m. The BS height is $8$m. The UE height is fixed at $1.5$m. Additional details of this 3GPP channel model are provided in Section~C-1 of \cite{Chatelier2023}. We use the 60 $\%$ clutter density scenario. This yields a highly NLoS environment (see \cite{Chatelier2023} for an illustrative example).

We consider two types of radio signal: the path gain (PG) and the channel impulse response (CIR) (both obtained via the statistical channel). The PG is obtained as the shadowing minus the path loss (in dB) (using the 3GPP terminologies \cite{TR38901}), see \cite{Chatelier2023} for additional details on the generation of these radio signals. The important point is that with the PG, one coefficient per BS is obtained whereas there are several coefficients per BS with the CIR. 
More specifically, for the PG datasets, for each position $k$, the PG at each of the 18 BS, say $\mathbf{a}_k \in \mathbb{R}^{1\times 18} $, is computed and the horizontal coordinate of the position $\mathbf{p}_k \in \mathbb{R}^{1 \times 2} $ is associated as the label. 
The CIR datasets are generated in a similar manner, but where the data for each position is now $\mathbf{b}_k \in \mathbb{C}^{256\times 18}$ instead of $\mathbf{a}_k$, where 256 is the length of the CIR. 
Spatial consistency, as specified in TR 38.901 \cite{TR38901}  (see Eq. 7.4-5 in TR 38.901) is implemented.
If a dataset with a reduced number of BS is required, we simply down-sample these datasets and keep a reduced number of columns per sample, where a column of a sample represent one BS.
Most of the simulations are performed using the PG datasets. Additional simulations using the CIR are provided in the Appendix.

We use the same neural networks as in \cite{Chatelier2023}:
For the PG datasets, a customized residual network (ResNet) architecture is used (see Fig. 4 in \cite{Chatelier2023} for a figure of the structure). 
Each fully connected layer is composed of $120$ neurons and the ReLU activation function is used. The neural network has 104.1k parameters.
For the CIR datasets, the ResNet18 architecture \cite{He2016} (with untrained weights) is considered. It has 11.1M parameters. This larger neural network is required due to the significantly larger size of the CIR radio signal.


For both neural networks used as NN1A, the output is an estimated horizontal coordinate vector $\hat{\mathbf{p}}_k$. 
The loss function for the training is the mean squared error, where the squared error is $\mathcal{L} = ||\hat{\mathbf{p}}_k - \mathbf{p}_k||^2$, where $\mathbf{p}_k$ is the true position.

For the case of the PG datasets, the above PG neural network is also used as NN1B. In this case, only the size of the input and output layers are modified accordingly. 
The rest of the architecture remains unchanged.  
The CIR neural network is not considered for NN1B: 
We did not perform simulations on the practical selection with the CIR datasets. Indeed, with the PG datasets, the output of the neural network NN1B remains of reasonable size (number of BS) and we can therefore use the same neural network architecture as NN1A. 
With the CIR, an output of size $256 \times number \ of  \ BS$  is required. It would therefore imply a more complex neural network. We leave this aspect for future work. 

\section{Simulation results}
\label{sec_sim_results}

\begin{table*}
\begin{center}
\begin{tabular}{||c | c | c | c | c|c| c|c||} 
 \hline
 $\mathcal{D}_1$ only & Rand $X=10\%$ & Rand $60\%$ & Rand $100\%$ &Genie test2 & Genie test1  &  App. genie test2 &  App. genie test1\\ [0.5ex] 
 \hline\hline
 0 & 6 & 30&40 & 20& 32 & 13 & 19\\ 
 \hline
\end{tabular}
\caption{Average performance gain (normalised $Q(0.9)$ value) with each training protocol for the PG datasets ($N=1700$).}
\label{table_perf_norma}
\end{center}
\end{table*}

\textbf{Simulation details.}
The baseline dataset contains 80 000 positions uniformly distributed in the scene.
The $N$ positions in $\mathcal{D}_1$ and the $N$ candidate positions are randomly sampled in the baseline dataset. 
Hence, we sample randomly chosen positions and compute the performance with each obtained dataset. This represents one realization.  
This is repeated 25 times for each tested value of the parameters. 
In other words, each point of the curves presented below is an average over 25 realizations.
In the simulations, $X\%=10\%$ of the $N$ candidate positions are selected, such that the size of $\mathcal{D}_2$ is 1.1 times the one of~$\mathcal{D}_1$.
The gain between the initial training (only with $\mathcal{D}_1$) and the additional training (with $\mathcal{D}_2$)  is computed as $G=$1- (pos. error addi. train.)/(pos. error ini. train). 

We consider two test datasets: 
A first test dataset, called \textbf{test1}, containing the non-selected positions among the $N$ candidate positions.
This represents a situation where all possible test positions are included in the candidate positions.
A second test dataset, called \textbf{test2}, containing the full 80000 positions except $\mathcal{D}_1$ and the selected candidate positions (i.e., except $\mathcal{D}_2$).
This corresponds to a case where only a subset of the test positions is included in the candidate positions.

Finally, the main evaluation metric is the 90$\%$ quantile of the cumulative distributive function (CDF) of the horizontal positioning
error $Q(0.9)$, i.e., the value $Q(0.9)$ (in meters) such that 90$\%$ of the errors are under $Q(0.9)$.

\begin{figure}
    \centering
\vspace{-10mm}
    \includegraphics[scale=.6]{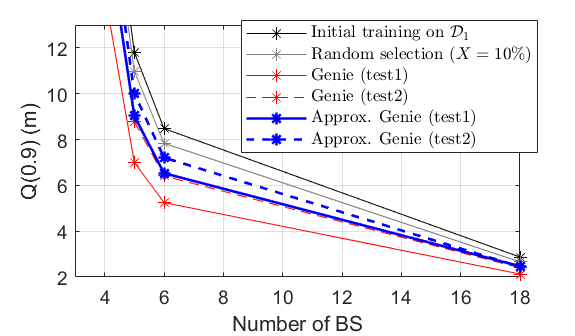}
    \caption{Performance of the genie approach and practical selection with the PG datasets ($N=1700$).}
    \label{fig:perf_PG_genie}
\vspace{-3mm}
\end{figure}

\textbf{Genie approach.}
We test the genie approach for data selection on the PG datasets with different number of BS. Fig.~\ref{fig:perf_PG_genie} shows the results. It depicts the performance of NN1A with the initial training only on $\mathcal{D}_1$ and the performance of NN2A after the additional training where the additional data is selected randomly (grey curves) and with the genie approach (red curves). Table~\ref{table_perf_norma}  also reports  the average\footnote{``Average" here is understood as averaging the gain of all points of one grey/red curve (with respect to the black curve).} normalized performance gain.

The average gain between the random selection (to obtain $\mathcal{D}_2$) and the initial training (with $\mathcal{D}_1$) is 6$\%$.
The average additional gain offered by the genie approach is between $14\%$ (test2 case) and $26\%$ (test1 case).

In terms of percentage of total achievable gain (i.e., using all $N$ candidate positions), Fig.~\ref{fig:perf_PG_genie} and Table~\ref{table_perf_norma} show that the genie approach enables to obtain more than 50$\%$ of this possible achievable gain instead of the $15\%$ with the random approach.

An alternative way to assess the performance is to quantify the amount of data saved for a fixed positioning performance.
Note that on Fig.~\ref{fig:perf_PG_genie} we included the performance of the random approach using $60\%$ of the $N$ candidate positions. This yields similar performance as the genie approach. Hence, we obtain (with a simple cross product) that the genie approach achieves similar performance as the random approach using approximately only 20$\%$ of the data.

Additional simulation results with the CIR datasets and for several values of $N$, to further assess the genie approach, are provided in the Appendix~\ref{simu_CIR_res}. 
These additional simulations confirm the trends observed with the PG datasets.

\begin{figure}[b]
\vspace{-7mm}
    \centering
    \includegraphics[scale=.43]{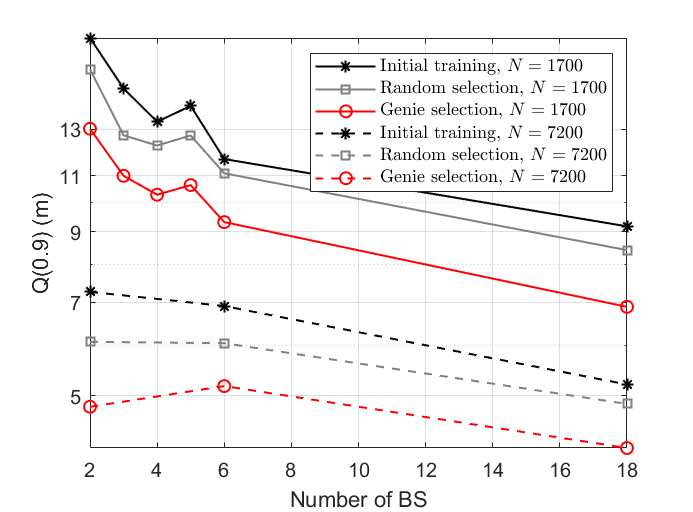}
    \caption{Performance of the genie approach with the CIR datasets.}
\vspace{-3mm}
    \label{fig:perf_CIR_genie}
\end{figure}

 Finally, we also briefly comment the performance only with $\mathcal{D}_1$, illustrated by the black curves (this does not involve the proposed data-selection method): 
Compared to \cite{Chatelier2023}, we let the training run for a longer time.  This enables to improve the performance compared to the one reported in \cite{Chatelier2023} (and get with the PG better performance than with the CIR\footnote{This not surprising as the CIR data is more complex as well as the required neural network architecture. As a result, the minimum required size of a dataset to get satisfactory performance is larger. This confirms the point made in \cite{Chatelier2023} that satisfactory performance can be obtained considering radio signals less complex than the CIR.} for $N=1700$, see also Fig.~\ref{fig:perf_CIR_genie}). 

\textbf{Practical approach.}
In this subsection, we provide the results with the proposed method to approximate the genie.
As a reminder, the key aspect of this method is the performance of NN1B, which should estimate the radio signal corresponding to a position.



The results are shown on Fig.~\ref{fig:perf_PG_genie} (two blue curves). 
The normalized performance gain of the practical selection is also reported in Table~\ref{table_perf_norma}.
We observe that the practical selection achieves approximately 50$\%$ of the gain achieved by the genie (the blue curves are equidistant of the grey and red curves). 

\textbf{Future work.}
The proposed technique is not the only manner to approximate the genie. 
There are probably other efficient techniques, which may be compared with this approach.
Note also that implementing this approach is more involved if the radio-signal size is high (such as the CIR): NN1B should then accurately output a high-dimensional vector.


\section{Conclusions}
In this paper, we investigated the problem of data selection to improve the training of deep learning models for positioning based on radio signals.
We first considered a genie approach to show that efficient data selection can bring significant performance gains compared to a random data selection.
Then, we proposed a practical algorithm to approximate the genie.
Simulation results show that both the genie and the practical approaches yield improved performance compared to the random data selection.
Alternatively, they enable to achieve the performance of the random data selection with a reduced dataset size.




\section{Appendix}
\label{simu_CIR_res}




We provide additional simulation results on the genie approach for data selection using the CIR datasets with different number of BS and with different values of $N$. They are reported in Fig.~\ref{fig:perf_CIR_genie}. 
Similarly to Fig.~\ref{fig:perf_PG_genie}, Fig.~\ref{fig:perf_CIR_genie} depicts the performance of NN1A with the initial training only on $\mathcal{D}_1$ and the performance of NN2A after the additional training where the additional data is selected randomly (grey curves) and with the genie approach (red curves). Only the case test1 is shown. 

These simulation results confirm the trends observed with the PG datasets. For each of the three sets of curves shown on the figures (two sets on Fig.~\ref{fig:perf_CIR_genie} and one on Fig.~\ref{fig:perf_PG_genie}), the average gain between the random selection and the initial training varies between $6\%$ and $9 \%$.
The average additional gain offered by the genie approach is approximately between $15-30\%$.

\vskip5pt

\noindent \textit{The authors are with Mitsubishi Electric Research and Development Centre Europe, 35700 Rennes, France}. 
\vskip3pt

\noindent E-mail: v.corlay@fr.merce.mee.com

\end{document}